\documentclass[a4paper,12pt,aip,rsi,superscriptaddress,reprint,floatfix]{revtex4-1}

\usepackage[utf8]{inputenc}
\usepackage[english]{babel}
\usepackage[T1]{fontenc}
\usepackage{amsmath,amssymb}
\usepackage{graphicx}
\usepackage{braket}
\usepackage{amsfonts}
\usepackage{array}
\usepackage{lmodern}
\usepackage{esvect}
\usepackage{fancyhdr}
\usepackage{xcolor}
\usepackage{textcomp}
\usepackage[squaren,Gray]{SIunits}
\usepackage{natbib}
\usepackage{hyperref}

\graphicspath{{Images/}}

\begin{document} 

\title{Two-color interferometer for the study of laser filamentation triggered electric discharges in air}

\author{Guillaume Point}
\email{guillaume.point@ensta-paristech.fr}
\affiliation{Laboratoire d'Optique Appliquée - ENSTA ParisTech, Ecole Polytechnique, CNRS - 828 boulevard des Maréchaux, 91762 Palaiseau, France}
\author{Yohann Brelet}
\affiliation{Laboratoire d'Optique Appliquée - ENSTA ParisTech, Ecole Polytechnique, CNRS - 828 boulevard des Maréchaux, 91762 Palaiseau, France}
\author{Leonid Arantchouk}
\affiliation{Laboratoire d'Optique Appliquée - ENSTA ParisTech, Ecole Polytechnique, CNRS - 828 boulevard des Maréchaux, 91762 Palaiseau, France}
\author{Jérôme Carbonnel}
\affiliation{Laboratoire d'Optique Appliquée - ENSTA ParisTech, Ecole Polytechnique, CNRS - 828 boulevard des Maréchaux, 91762 Palaiseau, France}
\author{Bernard Prade}
\affiliation{Laboratoire d'Optique Appliquée - ENSTA ParisTech, Ecole Polytechnique, CNRS - 828 boulevard des Maréchaux, 91762 Palaiseau, France}
\author{André Mysyrowicz}
\affiliation{Laboratoire d'Optique Appliquée - ENSTA ParisTech, Ecole Polytechnique, CNRS - 828 boulevard des Maréchaux, 91762 Palaiseau, France}
\author{Aurélien Houard}
\affiliation{Laboratoire d'Optique Appliquée - ENSTA ParisTech, Ecole Polytechnique, CNRS - 828 boulevard des Maréchaux, 91762 Palaiseau, France}

\begin{abstract}
We present a space and time resolved interferometric plasma diagnostic for use on plasmas where neutral-bound electron contribution to the refractive index cannot be neglected. By recording simultaneously the plasma optical index at 532 and \unit{1064}{\nano\metre}, we are able to extract independently the neutral and free electron density profiles. We report a phase resolution of \unit{30}{\milli\radian}, corresponding to a maximum resolution on the order of $\unit{4\times 10^{22}}{\rpcubic\metre}$ for the electron density, and of $\unit{10^{24}}{\rpcubic\metre}$ for the neutral density. The interferometer is demonstrated on centimeter-scale sparks triggered by laser filamentation in air with typical currents of a few tens of A.\\

Article published as Rev. Sci. Instrum. \textbf{85}, 123101 (2014) (\url{http://dx.doi.org//10.1063/1.4902533})

\end{abstract}

\maketitle

\section{Introduction}

Laser filamentation, that is the self-guided, non-diffracting propagation regime for an ultrashort laser pulse of which peak power is greater than a critical power (about \unit{5}{\giga\watt} at \unit{800}{\nano\metre}), results from the dynamic competition between the self-focusing Kerr effect, on the one hand, and diffraction, multiphoton absorption, and plasma defocusing on the other hand \cite{Couairon2007}. This phenomenon has many interesting features, among which is the fact that the laser pulse leaves a long underdense (n$_{e} \approx \unit{10^{22}}{\rpcubic\metre}$) plasma column in its trail \cite{Papazoglou2008,Chen2010a}. This column can be used for various applications, such as acting as a waveguide for infrared and microwave radiation \cite{Chateauneuf2008,Alshershby2013,Ren2013}, serving as a gain medium for lasing action \cite{Kartashov2012b,Point2014}, generating strong terahertz emission \cite{DAmico2007,Wang2011a}, or triggering and guiding electric discharges in atmospheric air \cite{Zhao1995,Rodriguez2002,Brelet2012a}. The latter is very promising due to its potential openings for the laser lightning rod \cite{Schwarz2003,Forestier2012}, high-voltage, high-current switches \cite{Rambo2001,Arantchouk2013}, or a virtual plasma antenna \cite{Dwyer1984,Brelet2012}.

The initiation, optimization and maturation of these technologies rely on a good knowledge of the discharge plasma parameters, that is the electron density and, to a lesser extent, the electron temperature. Interferometry had proven to be a very reliable and rigorous electron density diagnostic, able to give absolute  density measurements \cite{Hutchinson2002}. Therefore, it appears to be an ideal choice as a diagnostic for filamentation-guided electric discharges. 

However, a difficulty arises from the nature of studied sparks. Since energy deposition from the discharge in the medium leads to strong hydrodynamic effects \cite{Shneider2006}, and because the dynamics of studied plasmas are slower than the characteristic development time of such effects, two different contributions to the plasma optical index emerge. One of them comes from the free electron population. The other originates from the neutral-bound electrons, because the neutral density along the probe beam and the reference beam are not balanced. A way to separate these two contributions is to record the plasma optical index at two different wavelengths, so-called two-color interferometry  \cite{Alcock1966,David1967,Weber1997,Castro2006}. Using this technique, it becomes possible to isolate the free electron density, and to record neutral density as well. Two-color interferometry has also been used to make vibration-insensitive electron density measurements on large experiments, where such effects can become significant \cite{Lehecka1988,Kawahata1999,Tanaka2004}. This application is very similar to ours, consisting in discriminating between two different contributions to the phase shift. In the simpler case where the plasma optical index only depends on the free electron density, dual-wavelength interferometry can also be used to extend the available dynamic range, using a short wavelength to probe high-density areas, and shifting to a longer wavelength for low-density regions \cite{Sagisaka2006}.

In this Article, we introduce a two-color interferometer designed for the study of filamentation-triggered electric discharges, but that can however be readily adapted for use on any plasma where bound electron contribution to the refractive index cannot be neglected. Working principles of this kind of plasma diagnostic are described in section \ref{section2}. Interferogram processing from image recording to the extraction of radial density profiles is explained in section \ref{section3}. Finally, experimental results from the study of a centimeter-scale spark discharge triggered by a laser filament are presented in section \ref{section4}. From these results, we estimate the phase shift resolution of the interferometer to be at most \unit{30}{\milli\radian} for $\lambda = \unit{532}{\nano\metre}$. For a typical spark transverse size of \unit{200}{\micro\metre}, this gives minimum recoverable densities on the order of $\unit{4\times10^{22}}{\rpcubic\metre}$ for free electrons, and of $\unit{10^{24}}{\rpcubic\metre}$ for neutrals.

\section{Plasma two-color interferometry}
\label{section2}

\subsection{Theoretical aspects}

As said earlier, two distinct contributions arise in the plasma optical index: one comes from free electrons, and the other from electrons bound to neutrals and ions. The polarization density $\vv{P}$ can be written as:
\begin{equation}
\vv{P} = \vv{P}_{f} + \vv{P}_{b},
\end{equation}
where $\vv{P}_{f}$ corresponds to the free electronic dielectric response, and $\vv{P}_{b}$ to the bound electronic dielectric response. The free electronic polarization density can be found using the Drude model for a cold and non-magnetic plasma:
\begin{equation}
\vv{P}_{f} = -\epsilon_{0}\frac{\omega_{p}^{2}}{\omega^{2}\left(1-i\frac{2\pi\nu_{c}}{\omega}\right)}\vv{E},
\label{eq_electron_polarization}
\end{equation}
where $\epsilon_{0}$ is the vacuum permittivity, $\omega_{p} = \sqrt{n_{e}e^{2}/m_{e}\epsilon_{0}}$ is the plasma frequency ($n_{e}$, $e$ and $m_{e}$ being respectively the electron density, the electron charge and the electron mass), $\nu_{c}$ is the average collision frequency between free electrons and other species in the plasma, and $\omega$ denotes the angular frequency of the harmonic exciting laser electric field $\vv{E}$ \cite{}. In our case, the probe beam is either in the visible range, or in the near-infrared range, that is $\omega \sim \unit{\times10^{15}}{\rad\cdot\rp\second}$. As for the electron collision frequency, if we suppose that our plasma is dominated by electron-neutral collisions because of incomplete ionization, we can estimate this parameter as: $\nu_{c} \approx 6\times10^{-15}n_{n}[\mathrm{m}^{-3}]\mathrm{T}_{e}^{1/2}[\mathrm{eV}]\unit{}{\hertz} \sim \unit{1.5\times10^{13}}{\hertz}$ for $k_{B}$T$_{e} \sim \unit{1}{\electronvolt}$ \cite{Rax2005}. We are thus in the limit $\omega \gg \nu_{c}$. Therefore, it is possible to simplify equation \eqref{eq_electron_polarization}:
\begin{equation}
\vv{P}_{f} = -\epsilon_{0}\frac{\omega_{p}^{2}}{\omega^{2}}\vv{E} = -\epsilon_{0}\frac{r_{e}\lambda^{2}}{\pi}n_{e}\vv{E},
\end{equation}
where the classical electron radius $r_{e} = e^{2}/4\pi\epsilon_{0}m_{e}c^{2}$ and the wavelength $\lambda = 2\pi c/\omega$ were introduced.

As for the bound electronic polarization density, it can be evaluated considering bound electrons as harmonic oscillators:
\begin{equation}
\vv{P}_{b} = \frac{e^{2}}{m_{e}}\left(n_{i}\sum_{k_{1}}\frac{f_{k_{1}}}{\omega_{0,k_{1}}^{2}-\omega^{2}}+n_{n}\sum_{k_{2}}\frac{f_{k_{2}}}{\omega_{0,k_{2}}^{2}-\omega^{2}}\right)\vv{E},
\label{eq_neutral_polarization}
\end{equation}
where $f_{k}$ is the oscillator strength associated with the $k$th resonance of which characteristic frequency is $\omega_{k}$, $n_{i}$ is the ion density, and $n_{n}$ the neutral density. At this point, the first term of the left-hand side can be neglected for two reasons: first, the studied sparks are only moderately ionized, meaning $n_{n} \gg n_{i}$. Second, ionic polarizability, proportional to the left-hand sum in equation \eqref{eq_neutral_polarization}, is less important than neutral polarizability \cite{Castro2006}. Consequently, we can write:
\begin{equation}
\begin{array}{ll}
\vv{P}_{b} &= \displaystyle\frac{e^{2}}{m_{e}}n_{n}\sum_{k}f_{k}\frac{1}{\omega_{0,k}^{2}-\omega^{2}}\vv{E}\\
 &= \displaystyle\epsilon_{0}\frac{r_{e}}{\pi}n_{n}\sum_{k}f_{k}\frac{\lambda^{2}\lambda_{o,k}^{2}}{\lambda^{2}-\lambda_{0,k}^{2}}\vv{E}.
\end{array}
\end{equation}

Since our probe laser has a very weak intensity and because the spark plasma is not magnetized, it can be locally considered as linear and isotropic. Thus we have:
\begin{equation}
\vv{P} = \epsilon_{0}(\epsilon_{r}-1)\vv{E},
\end{equation}
with the relative dielectric constant $\epsilon_{r}$ being linked to the plasma optical index $n_{p}$ following:
\begin{equation}
\epsilon_{r} = n_{p}^{2}.
\end{equation}
Therefore, the plasma index can be written as:
\begin{equation}
\begin{array}{ll}
n_{p} &= \displaystyle \sqrt{1-\frac{r_{e}\lambda^{2}}{\pi}n_{e}+\frac{r_{e}}{\pi}n_{n}\sum_{k}f_{k}\frac{\lambda^{2}\lambda_{o,k}^{2}}{\lambda^{2}-\lambda_{0,k}^{2}}} \\
& \approx \displaystyle 1-\frac{r_{e}\lambda^{2}}{2\pi}n_{e}+\frac{r_{e}}{2\pi}n_{n}\sum_{k}f_{k}\frac{\lambda^{2}\lambda_{o,k}^{2}}{\lambda^{2}-\lambda_{0,k}^{2}},
\end{array}
\label{eq_plasma_index}
\end{equation}
$|n_{p} - 1|$ being on the order of $10^{-4}$. The last term in this last equation is actually nearly independent of the wavelength in the visible-near infrared range, and is usually written as:
\begin{equation}
\frac{r_{e}}{2\pi}n_{n}\sum_{k}f_{k}\frac{\lambda^{2}\lambda_{o,k}^{2}}{\lambda^{2}-\lambda_{0,k}^{2}} = \frac{\beta}{n_{0}}n_{n},
\end{equation} 
called the Gladstone-Dale relation. The Gladstone-Dale constant $\beta$ defines the value of the refractive index $1+\beta$ at a reference point characterized by $n_{n} = n_{0}$. Data computed using the Ciddor equation \cite{Ciddor1996} with standard air  temperature and pressure condition give $\beta \approx 2.7 \times10^{-4}$ for a reference neutral density $n_{0} = \unit{2.47 \times 10^{25}}{\rpcubic\metre}$. The plasma index is now reduced to:
\begin{equation}
n_{p} = 1-\frac{r_{e}\lambda^{2}}{2\pi}n_{e}+\frac{\beta}{n_{0}}n_{n}.
\end{equation}

When an interferometric measurement of the plasma in air is done, the recorded phase shift of the fringes $\Delta\varphi$ is linked to the plasma refractive index by:
\begin{equation}
\Delta\varphi(\lambda) = \frac{2\pi}{\lambda}\int_{s_{1}}^{s_{2}}(n_{p}(\lambda)-1-\beta)~\mathrm{d}s,
\end{equation}
$s_{1}$ and $s_{2}$ defining the length of integration chord. If one records simultaneously the phase shift due to the plasma at two different wavelengths along the same chord, one can then separate the contributions from free and bound electrons in the plasma index as follows:
\begin{equation}
\left\{
\begin{array}{ll}
& \displaystyle \int_{s_{1}}^{s_{2}}n_{e}(s)~\mathrm{d}s = \frac{\lambda_{1}\Delta\varphi_{1}-\lambda_{2}\Delta\varphi_{2}}{r_{e}(\lambda_{2}^{2}-\lambda_{1}^{2})}\\
& \displaystyle \int_{s_{1}}^{s_{2}}(n_{n}(s)-n_{o})~\mathrm{d}s = \frac{n_{o}}{2\pi\beta}\frac{\lambda_{2}\Delta\varphi_{1}-\lambda_{1}\Delta\varphi_{2}}{\frac{\lambda_{2}}{\lambda_{1}}-\frac{\lambda_{1}}{\lambda_{2}}}.
\end{array}
\right.
\label{eq_line_densities}
\end{equation}

\subsection{Experimental setup}

\begin{figure}[!ht]
\begin{center}
\includegraphics[width = 7.9cm]{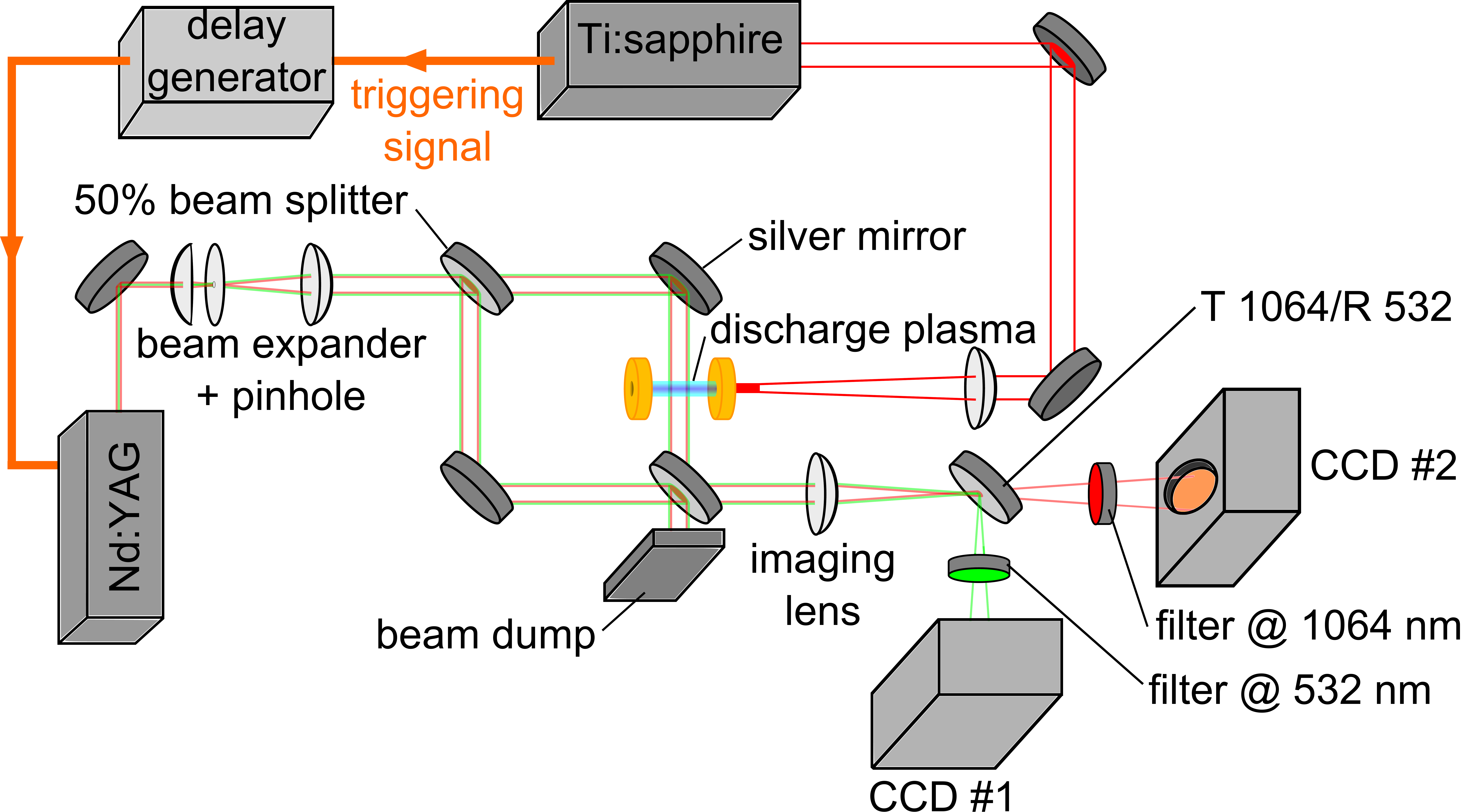}
\end{center}
\caption{Schematic layout of the interferometric setup.}
\label{interferometer}
\end{figure}

Our interferometer is built in a standard Mach-Zehnder configuration (cf. figure \ref{interferometer}). The probe laser (Quanta Ray GCR-290-10 from Spectra Physics) is a Nd:YAG Q-switched, frequency doubled pulsed laser delivering $\sim \unit{100}{\micro\joule}$, \unit{8}{\nano\second} full width at half maximum (FWHM) light pulses at both 532 and \unit{1064}{\nano\metre}. The probe beam is spatially cleaned and magnified using a beam expander and a \unit{100}{\micro\metre} pinhole, yielding a pseudo-Gaussian collimated beam with a \unit{7.2}{\milli\metre} FWHM at \unit{532}{\nano\metre} and \unit{8.3}{\milli\metre} FWHM at \unit{1064}{\nano\metre}. The delay between the two probe pulses due to the propagation through optical materials is evaluated to a few ps, which is negligible with respect to the duration of laser pulses. The whole interferometer is built on a $\unit{60\times60}{\centi\metre\squared}$ breadboard mounted on mechanic isolation feet to reduce vibrations, and entirely enclosed to limit the influence of air turbulence on measurements.

An ultrashort laser pulse from the Ti:sapphire chirped pulse amplification laser chain ENSTAmobile is focused at $f/35$ and collapses into a filament in the discharge gap, triggering the studied spark. The resulting discharge plasma is generated in one of the arms of the interferometer, perpendicularly to the probe beam. The probe laser is synchronized with the ENSTAmobile using the latter's internal clock, enabling us to precisely adjust the temporal delay between the laser pulses between 0 and \unit{100}{\milli\second} with a ns precision and a \unit{1.5}{\nano\second} jitter.

Interferograms are recorded by means of two CCD cameras (TaperCamD-UCD12 from DataRay, Inc.) with a $1360 \times 1024$ pixel matrix and a \unit{10.5}{\micro\metre} pixel size. Probe wavelengths are separated using a dichroic mirror (T1064/R532). Adapted bandpass filters with $< \unit{10}{\nano\metre}$ bandwidth suppress remaining broadband emission from the plasma. The whole detection part is arranged in a $2f/2f$ fashion using a \unit{75}{\milli\metre} focusing lens, CCD cameras being in the conjugated plane of the plasma plane.

The interferometer is slightly misaligned so that $\sim 10/20$ fringes (depending on the wavelength) appear on the detectors in absence of plasma in the probing arm.

\section{Interferogram processing}
\label{section3}

\subsection{Phase recovery algorithm}

The first step to perform during the processing of interferograms is to retrieve relative phase planes from images. In our case, this is done using a 1D continuous wavelet transform (CWT) algorithm. This kind of algorithm has been proved to be more reliable than routines based on the fast Fourier transform (FFT), particularly in case of low contrast and/or noisy interferograms \cite{Tomassini2001,Liu2004}. We will first describe the CWT, and then how this mathematical transform is used to extract the local phase from interferograms. 

\subsubsection{The continuous wavelet transform}

The CWT is a mathematical integral transform for any square-integrable function $f \in L_{2}(\mathbb{R})$, much like the Fourier transform. However, where the Fourier transform decomposes $f$ on a complete basis of trigonometric polynomials, the CWT projects $f$ on a base of particular functions, called wavelets, which are well localized both in direct and Fourier space. If we consider a function $\psi \in L_{2}(\mathbb{R})$, called the mother wavelet, satisfying particular conditions (zero average, normalized energy, and a last condition called admissibility condition \cite{Mallat2009}), then the wavelet transform of $f$ is defined as:
\begin{equation}
f_{CWT}(a,b) = \braket{f|\psi_{a,b}} = \frac{1}{\sqrt{a}}\int_{\mathbb{R}}f(x)\psi\left(\frac{x-b}{a}\right)^{*}~\mathrm{d}x,
\end{equation}
where $\psi_{a,b}$ is the daughter wavelet of parameters $a$ and $b$, $a \in \mathbb{R}^{*}_{+}$ is called the scale parameter, $b \in \mathbb{R}$ is the translation parameter, and $*$ corresponds to the complex conjugation operation. The CWT thus projects a 1D function on a 2D space of which $(a,b)$ forms a suitable basis, called time/frequency space. This is because $b$ is dimensionally similar to $x$ (that is, the time direction) while $a$ defines the daughter wavelet extension in the frequency space, and can be directly linked to the Fourier frequency \cite{Meyers1993}.

The mother wavelet we use is called the Morlet wavelet, which is a trigonometric polynomial modulated by a Gaussian envelope:
\begin{equation}
\forall x \in \mathbb{R}, \psi^{M}_{s}(x) = \frac{1}{\sqrt[4]{\pi}}\mathrm{e}^{-x^{2}/2}\mathrm{e}^{isx},
\end{equation}
where $s \ge 6$ to satisfy the admissibility condition \cite{Mallat2009}. This wavelet is particularly well adapted for the study of sinusoidal signals because of its good localization in both direct and frequency space \cite{Gdeisat2006}.

\begin{figure}[!ht]
\begin{center}
\includegraphics[width = 7.9cm]{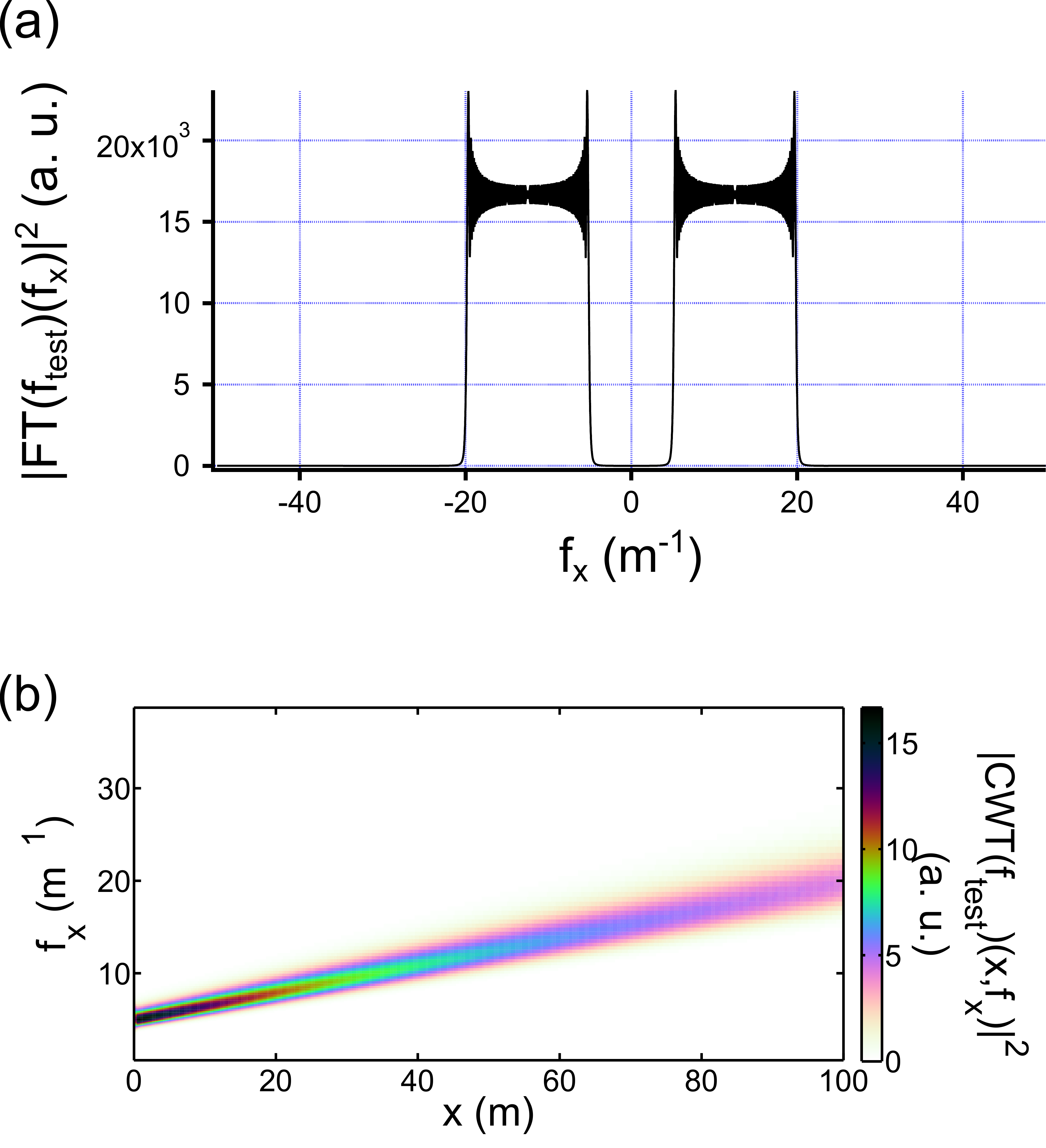}
\end{center}
\caption{Comparison between the Fourier transform and the CWT. (a): squared magnitude of the Fourier transform of $f^{test}$. (b): squared magnitude of the CWT of $f^{test}$.}
\label{chirp}
\end{figure}

The power of the CWT over Fourier transform becomes evident when it comes to the study of signals of which frequency varies with time. If we take the example of a linearly chirped cosine function:
\begin{equation}
f_{test}(x) = \cos\left(2\pi\left(5+0.075x\right)x\right),
\end{equation}
we can compute both the Fourier transform and the CWT using a Morlet wavelet for this function. As seen on figure \ref{chirp}-(a), the spectral power density for $f_{test}$ calculated using a standard FFT algorithm is different from 0 only between 5 and \unit{20}{\rp\metre}, which corresponds to the signal frequency $f_{x}$ at $x = 0$ and $x = \unit{100}{\metre}$, i. e. the space boundaries for the study. However, no additional information can be retrieved from this plot.

The CWT, however, as a time/frequency analysis tool, gives much more information about the studied signal. Thus, as shown in figure \ref{chirp}-(b), it is possible to retrieve the frequency of this signal for any $x$ value using the curve of maximum magnitude, that is the ridge of the wavelet transform. The ridge is particularly significant in wavelet analysis because along this curve, the phase of $f_{CWT}$ is equal to the phase of $f$ \cite{Mallat2009}.

\subsubsection{The algorithm}

\begin{figure}[!ht]
\begin{center}
\includegraphics[width = 7cm]{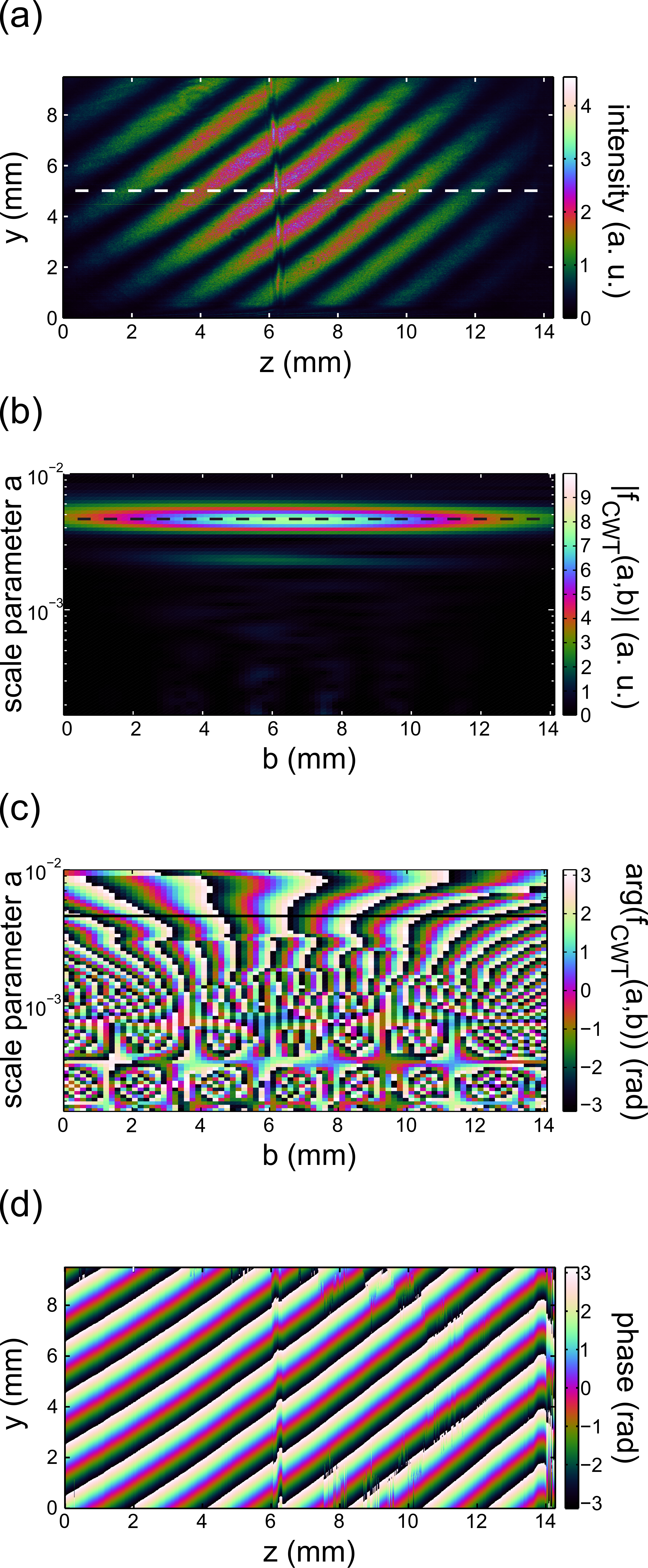}
\end{center}
\caption{Principle of the phase recovery algorithm. (a): example of recorded interferogram. The test row is indicated by the white dashed line. (b): CWT magnitude for the test line. The ridge is indicated by the black dashed line. (c): CWT phase for the test line. The ridge position is reported here and indicated by the black solid line. (d): phase recovered from the interferogram.}
\label{phase_algorithm}
\end{figure}

The phase recovery algorithm works as follows: each row of the interferogram is selected one after the other, as illustrated in figure \ref{phase_algorithm}-(a). For a given line, the CWT is computed by means of a routine based on the one proposed by Torrence and Compo using a Morlet wavelet as mother wavelet \cite{Torrence1998}. This yields a 2D complex array, which can be split into a magnitude and a phase plane (figure \ref{phase_algorithm}-(b) and (c), respectively). The following step is to recover the ridge position on the magnitude plane. Because of noise and low interferogram contrast, the true ridge corresponding to the fringe frequency can be dwarfed by parasite curves on the wavelet magnitude map. 

A way to overcome this and to enhance the robustness of the algorithm with respect to noise is to add a criterion for the ridge selection other than looking for the curve of maximum modulus. A choice is to discard candidate ridges presenting too pronounced steps along $a$, since noise tends to generate parasite curves that often shift in frequency. Thus, to recover the correct ridge using this procedure, one must look for the parametric curve ($a_{0} = g(b)$) that minimizes a cost function defined as:
\begin{equation}
\begin{array}{ll}
\mathrm{cost}(g(b),b) =  &-A_{1}\displaystyle\int_{-\infty}^{b}|f_{CWT}(g(b'),b')|^{2}~\mathrm{d}b'\\
&\displaystyle + A_{2}\int_{-\infty}^{b}\left|\frac{\partial g}{\partial b'}(b')\right|^{2}~\mathrm{d}b',
\end{array}
\end{equation}
where $A_{1}$ and $A_{2}$ are weighting coefficients \cite{Liu2004}. For these parameters, we typically use values such as $\displaystyle A_{2}/A_{1} \approx \max_{a,b}|f_{CWT}(a,b)|^{2}$.

Once the ridge position is recovered, it becomes possible to find back the phase of the original signal using the phase of the CWT since on the ridge, both values are equal. Reporting the ridge position on the phase map thus yields directly the phase along the studied line, as illustrated on figure \ref{phase_algorithm}-(c). By repeating this operation for every line of the interferogram, we can finally recover the phase for the whole image (figure \ref{phase_algorithm}-(d)).

\subsection{Phase unwrapping algorithm}

As seen in figure \ref{phase_algorithm}-(d), the phase recovery algorithm extracts \textit{wrapped} phase, that is a phase map presenting $2\pi$ discontinuities. This is because the arctangent function used to evaluate the phase of the wavelet transform returns its principal value in the $[-\pi,\pi[$ interval. The following step therefore consists in unwrapping the extracted phase. The most basic algorithm to perform this task simply follows a predefined path, usually one row/column after the other, and looks for phase jumps greater than $\pi$ between two adjacent pixels, correcting the phase of  untreated pixels by adding or subtracting $2\pi$ multiples. Of course, this approach is very basic and suffers from several problems: first, noisy phase map and low signal-to-noise ratio can trick the algorithm and result in the appearance of artificial phase discontinuities. Second, any error along the unwrapping path can be propagated and corrupt the whole phase map.

For this reason, we make use of an unwrapping algorithm that follows a non-continuous path defined by decreasing value of a reliability function calculated for every pixel on the phase map. In our case, this reliability function depends on the second difference between adjacent pixels, that is the squared phase difference between a given pixel and its two neighbors along the same direction, corrected, if necessary, from any $2\pi$ discontinuity. For example, considering the pixel $(i,j)$, the horizontal second difference for this pixel, $H^{2}(i,j)$, would be:
\begin{equation}
H^{2}(i,j) = (w(\varphi(i,j-1)-\varphi(i,j))-w(\varphi(i,j)-\varphi(i,j+1)))^{2},
\end{equation}
where $w$ represents the wrapping operator. Computing vertical $V^{2}(i,j)$ and diagonal $D_{1}^{2}(i,j)$ and $D_{2}^{2}(i,j)$ second differences allows to deduce the reliability function for the pixel $(i,j)$, $R(i,j)$, as:
\begin{equation}
R(i,j) = (H^{2}(i,j)+V^{2}(i,j)+D_{1}^{2}(i,j)+D_{2}^{2}(i,j))^{-1/2}.
\end{equation}
The unwrapping operation works along the path going from the most reliable pixel to the least reliable one, joining processed pixels into groups, which can be unwrapped one with respect to the other as a whole. This algorithm was first proposed and demonstrated by Herr\'{a}ez \textit{et al.} \cite{Herraez2002}, both robust and fast.

\begin{figure}[!ht]
\begin{center}
\includegraphics[width = 7.9cm]{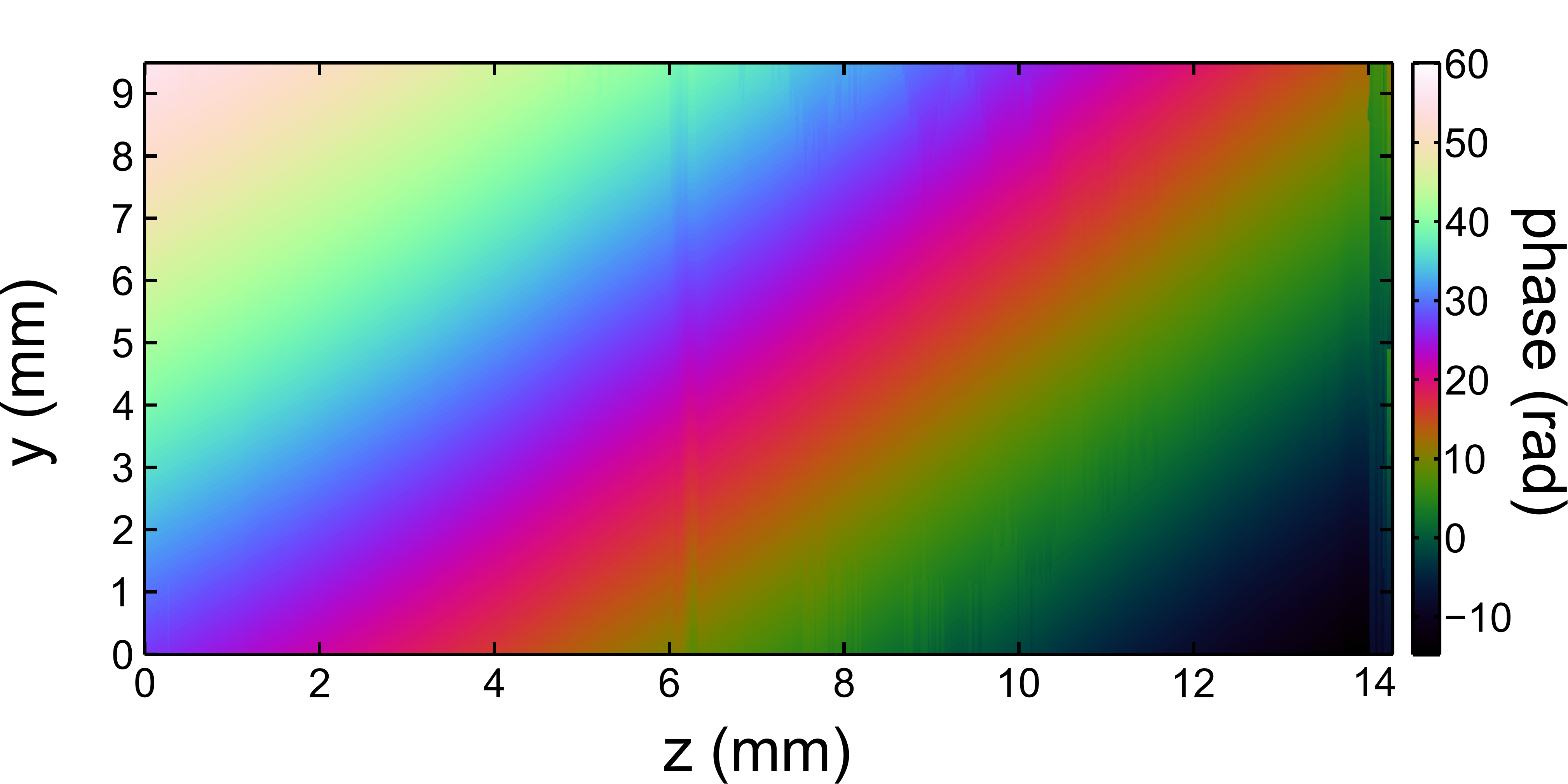}
\end{center}
\caption{Unwrapped phase map obtained after running the unwrapping algorithm on the wrapped phase map of figure \ref{phase_algorithm}-(d).}
\label{fig_phase_unwrapping}
\end{figure}

Figure \ref{fig_phase_unwrapping} presents the phase map resulting from the unwrapping algorithm applied to the wrapped phase map displayed in figure \ref{phase_algorithm}-(d). One can see that no unwrapping error, that is artificial phase jump, was introduced, and that phase noise initially present on the wrapped map is still confined to the same areas. Now that unwrapped phase maps can be reliably extracted from interferograms, it becomes possible to compute phase shift maps.

\subsection{Phase shift extraction}

Extracting useful information from interferograms requires the evaluation of the phase shift, that is the removal of the background phase due to the undisturbed spatial carrier. This step is done by recording several blank interferograms (i. e. without plasma) and recovering the corresponding unwrapped phase map, which includes any wavefront distortion induced by optical components in the interferometer. For any interferogram recorded in presence of plasma, a simple algorithm looks for the reference phase plane minimizing the function $\sigma(\varphi)$ defined as:
\begin{equation}
\sigma(\varphi) = \sqrt{\left\langle(\varphi-\varphi_{ref})^{2}\right\rangle-\left(\left\langle\varphi-\varphi_{ref}\right\rangle\right)^{2}},
\end{equation}
i. e. the standard deviation of the difference between the  phase plane and the reference candidate phase plane evaluated on the whole image. 

\begin{figure}[!ht]
\begin{center}
\includegraphics[width = 7.9cm]{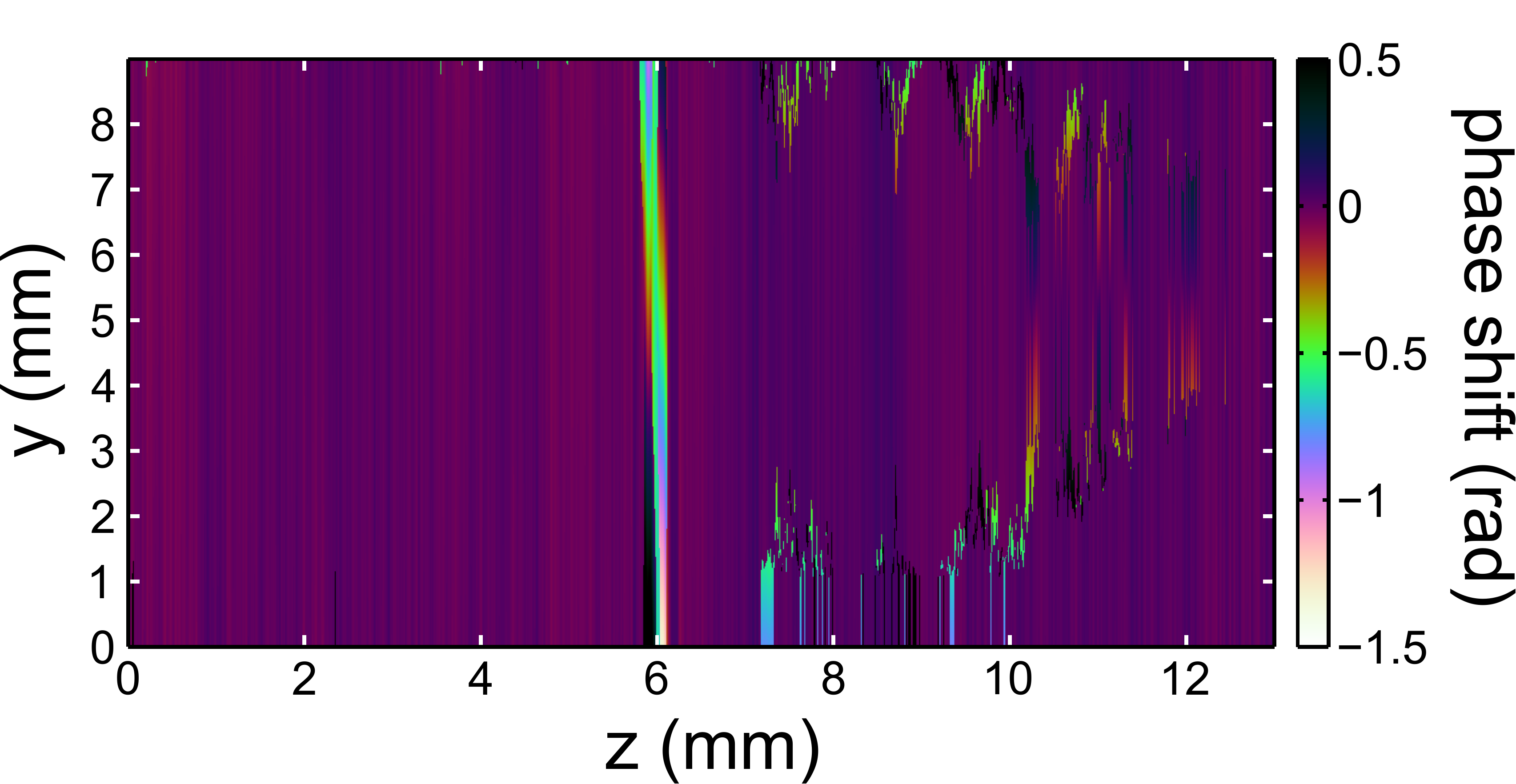}
\end{center}
\caption{Phase shift plane evaluated from the interferogram displayed in figure \ref{phase_algorithm}-(a).}
\label{fig_phase_shift}
\end{figure}

Once the best reference candidate is found, it is subtracted from the informative phase plane. Usually, a phase offset persists after this step. To correct this, the average phase outside the plasma-affected region is computed and removed from the whole phase map, giving a globally null phase shift out of the plasma. For instance, the interferogram displayed in figure \ref{phase_algorithm}-(a) results in the phase shift map of figure \ref{fig_phase_shift} using an appropriate reference phase map. Phase noise can also be evaluated as the phase standard deviation over the undisturbed region. We typically find a limiting value of \unit{30}{\milli\radian} RMS at \unit{532}{\nano\metre}, which results, for a probed object of size $\sim\unit{200}{\micro\metre}$, in a density resolution of $\unit{4\times10^{22}}{\rpcubic\metre}$ for free electrons and of $\unit{10^{24}}{\rpcubic\metre}$ for neutrals.

Phase shift maps can then be used directly to evaluate line-integrated densities following equation system \eqref{eq_line_densities}.

\subsection{Abel inversion algorithm}

\subsubsection{Principle}

\begin{figure}[!ht]
\begin{center}
\includegraphics[width = 6.5cm]{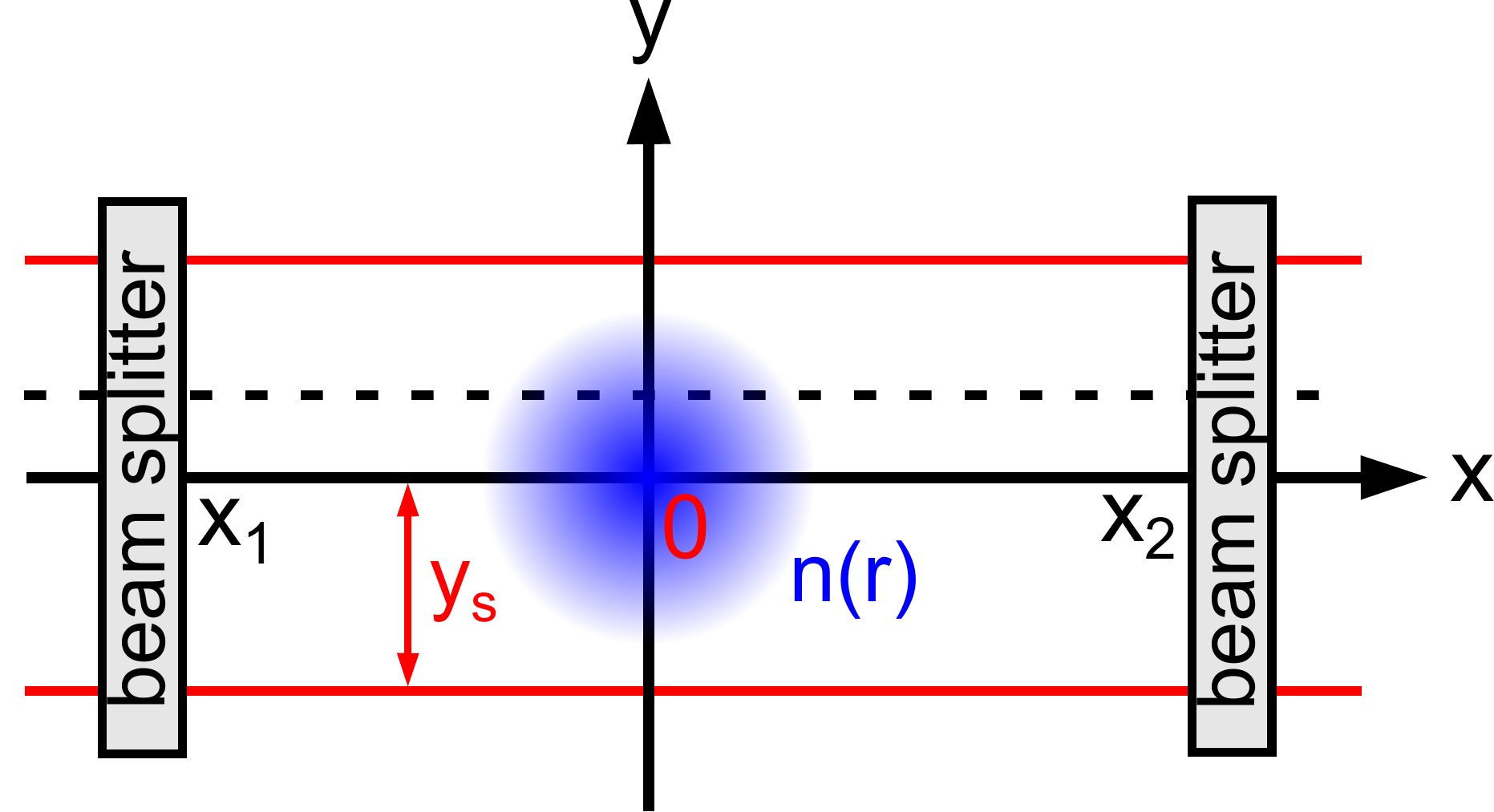}
\end{center}
\caption{Schematic depiction of the probe arm of the interferometer for the Abel inversion problem.}
\label{fig_Abel_inversion}
\end{figure}

As given by equation \eqref{eq_line_densities}, only line-integrated densities along the probe beam propagation direction are recovered from interferograms. The very last step therefore consists of recovering true densities. In the general case where no assumption is made about the density function, this inversion problem has no solution. A way to approach a solution is to multiply probing chords in different directions, that is a tomographic reconstruction \cite{Hutchinson2002}. However, in the case of cylindrically symmetrical density functions, a rigorous mathematical solution exists. This is particularly important in the field of plasma diagnostic because many laboratory plasmas, such as ours, exhibit this kind of symmetry. Considering the schematic representation for the Mach-Zehnder interferometer displayed in figure \ref{fig_Abel_inversion}, the recorded information at the output is:
\begin{equation}
f(y,z) = \int_{x_{1}}^{x_{2}}\rho(x,y,z)~\mathrm{d}x,
\end{equation}
where $\rho$ represents either free electron density of relative neutral density. Assuming that $\rho$ is cylindrically symmetrical, we can rewrite this equation:
\begin{equation}
f(y,z) = \displaystyle \int_{y}^{\sqrt{x_{1}^{2}+y^{2}}}\frac{r\rho(r,z)}{\sqrt{r^{2}-y^{2}}}~\mathrm{d}r +\displaystyle \int_{y}^{\sqrt{x_{2}^{2}+y^{2}}}\frac{r\rho(r,z)}{\sqrt{r^{2}-y^{2}}}~\mathrm{d}r.
\end{equation}

If $x_{1}$ and $x_{2}$ are large enough with respect to the plasma characteristic dimension, that is if $\forall (y,z), \rho(x_{1},y,z) \approx \rho(x_{1},y,z) \approx 0$ , then:
\begin{equation}
f(y,z) = 2\int_{y}^{+\infty}\frac{r\rho(r,z)}{\sqrt{r^{2}-y^{2}}}~\mathrm{d}r,
\end{equation}
which is called the Abel transform of the function $\rho$. At the condition that $\forall z, \displaystyle\lim_{r \rightarrow + \infty}r\rho(r,z) = 0$, we can invert the Abel transform and extract back $\rho$ from $f$ following:
\begin{equation}
\rho(r,z) = - \frac{1}{\pi} \int_{r}^{+\infty}\frac{\partial f}{\partial y}(y,z)\frac{\mathrm{d}y}{\sqrt{y^{2}-r^{2}}}.
\label{eq_inverse_Abel}
\end{equation}
This last step can only be performed if the integral converges, i. e. if the probe spatial extension $y_{s}$ is large enough so that $\forall z, f(\pm y_{s},z) \approx 0$.

Using this last equation, we are now able to recover true densities $n_{e}(r,z)$ and $n_{n}(r,z)$ from two-color interferograms.

\subsubsection{The algorithm}

Several methods can be used to numerically implement the inverse Abel transform. Obviously, the most direct approach is to discretize equation \eqref{eq_inverse_Abel}. However, two major difficulties arise here. The first one lies in the spatial derivative term which intrinsically amplifies data noise. The other one is the pole $y = r$ in the integral.

A way to avoid these problems is to take advantage of the fact that the Abel transform is a component of the FHA cycle, namely is equivalent to a composition of an inverse Fourier transform and of a Hankel transform of order 0. Thus, equation \eqref{eq_inverse_Abel} can be rewritten as:
\begin{equation}
\rho(r,z) = \frac{1}{2\pi}\int_{0}^{+\infty}\left(\int_{-\infty}^{+\infty}f(y,z)\mathrm{e}^{-iky}~\mathrm{d}y\right)kJ_{0}(kr)~\mathrm{d}k,
\end{equation}
where $J_{0}$ is the 0-order Bessel function of the first kind. In this last equation, the spatial derivative and the pole have disappeared, so it can be used for discretization following:
\begin{equation}
\rho(r_{i},z_{j}) = \frac{\Delta k\Delta y}{2\pi}\sum_{l=0}^{N}k_{l}J_{0}(r_{i}k_{l})\sum_{m=-N}^{N}f(y_{m},z_{j})\cos(k_{l}y_{m}),
\end{equation}
where it has been taken into account that $f$ is an even function. This kind of algorithm is called Fourier-Hankel, and is particularly fast because of the FFT \cite{Smith1988,Alvarez2002}.

\begin{figure}[!ht]
\begin{center}
\includegraphics[width = 7cm]{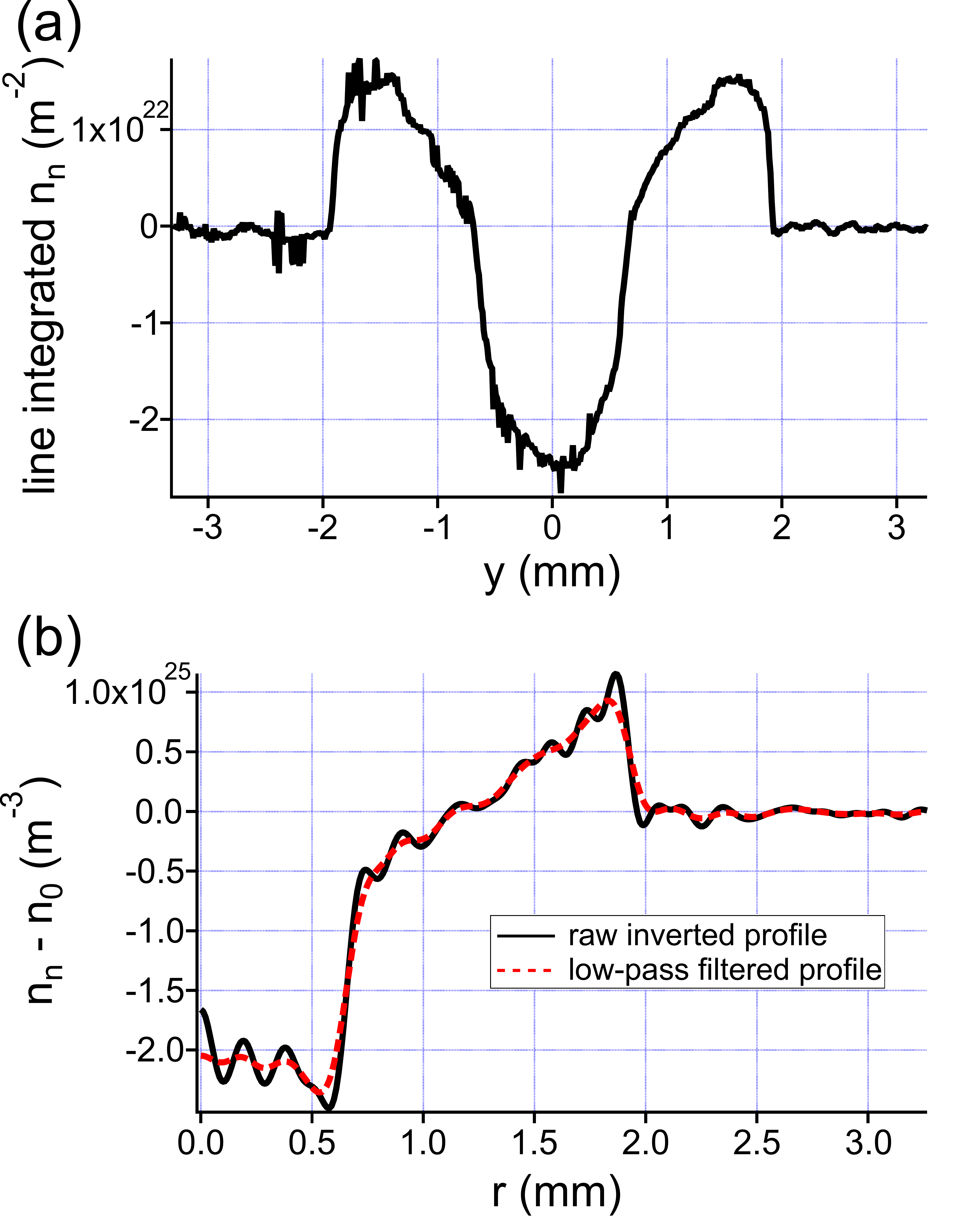}
\end{center}
\caption{(a): example of line-integrated neutral density profile. (b): corresponding inverted neutral density profiles using raw experimental data (black solid curve) or using a low-pass filter during the inversion process (red dashed curve).}
\label{fig_Abel_algo}
\end{figure}

Another useful feature arising from the Fourier-Hankel algorithm lies in the possibility to implement a low-pass filtering of experimental data during the process. Indeed, the first step of the transform involves a Fourier transform. At this point, one can then convolute the corresponding spectrum with an appropriate filtering window (in our case, a Hann window), before implementing the inverse Hankel transform. The usefulness of this technique is illustrated in figure \ref{fig_Abel_algo}. Figure \ref{fig_Abel_algo}-(a) gives an example of integrated neutral density profile, while figure \ref{fig_Abel_algo}-(b) shows the corresponding inverted profiles using the Fourier-Hankel algorithm. In the first case, only raw experimental data are used, resulting in few unrealistic oscillations (black solid curve). Using the low-pass filtering step yields a much smoother output profile (red dashed curve).

\section{Experimental results}
\label{section4}

\subsection{Setup for electric discharges}

\begin{figure}[!ht]
\begin{center}
\includegraphics[width = 6cm]{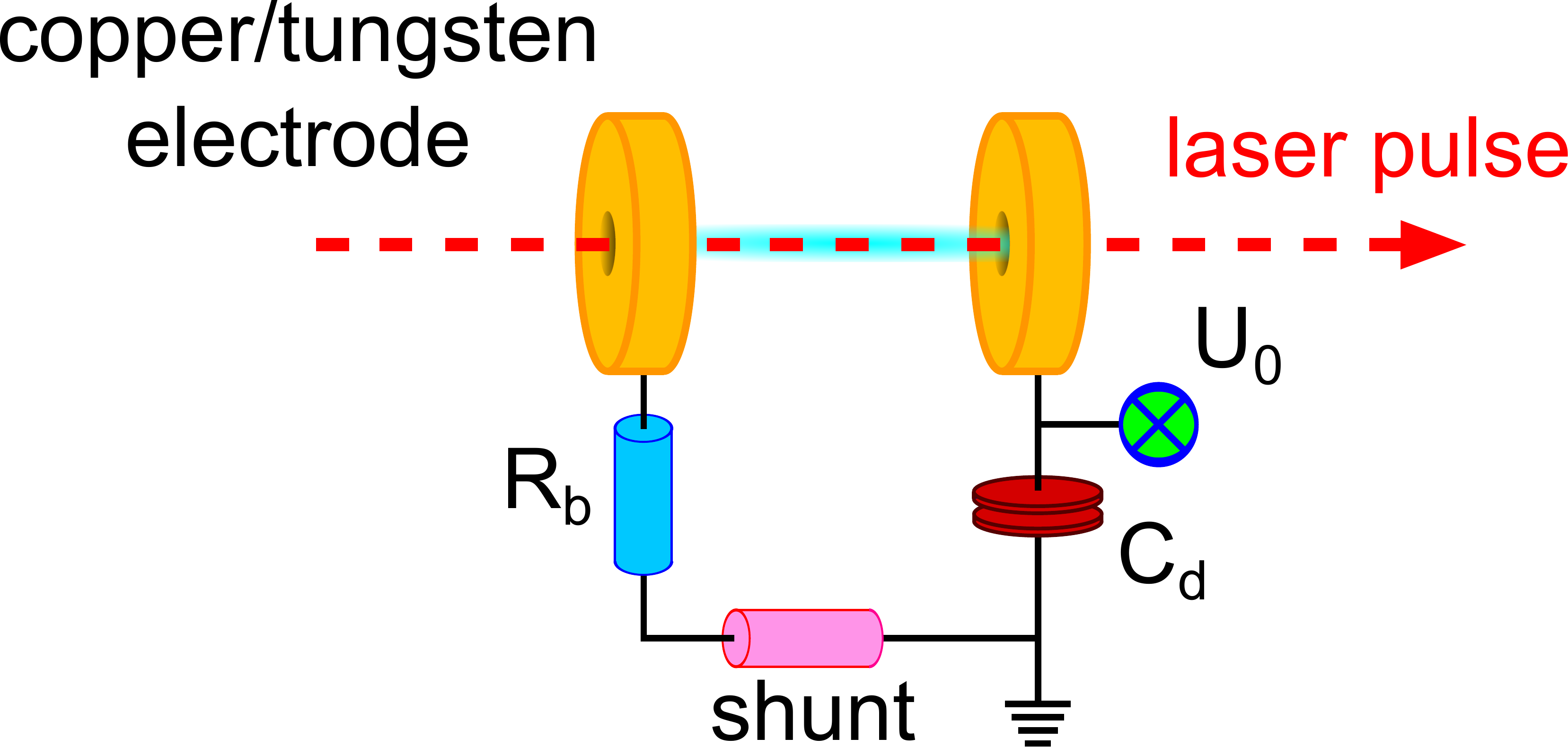}
\end{center}
\caption{Schematic depiction of the electrical circuit.}
\label{electric_circuit}
\end{figure}

Figure \ref{electric_circuit} gives a schematic overview of the experimental setup. A storage capacitor of nominal value C$_{d} = \unit{2}{\nano\farad}$ charged up to a voltage U$_{0} = \unit{15}{\kilo\volt}$ DC discharges through a laser triggered spark gap to a ballast resistance of nominal value R$_{b} = \unit{380}{\ohm}$ and a current viewing resistor, or shunt, of \unit{50}{\milli\ohm}.


The gap switch consists in two cylindrical copper-tungsten alloy electrodes of face diameter \unit{36}{\milli\metre} placed into a Plexiglas construction opened to atmospheric air \cite{Arantchouk2013}. The laser beam passes through \O \unit{3}{\milli\metre} holes drilled into the electrodes. The gap spacing between the electrodes is \unit{10}{\milli\metre} corresponding to a self-breakdown level U$_{cr} \sim \unit{30}{\kilo\volt}$. Working at U$_{0} = \unit{15}{\kilo\volt}$ thus ensures that no breakdown can occur without external triggering.

The discharge is triggered with a femtosecond laser pulse (\unit{4.5}{\milli\joule}, \unit{50}{\femto\second} @ \unit{800}{\nano\metre}) collapsing to form a single filament about \unit{2}{\centi\metre} long. Discharge onset results in a current pulse with a rise time of about \unit{18}{\nano\second} decaying exponentially with a damping time R$_{b}$C$_{d} \sim \unit{800}{\nano\second}$. Its maximal amplitude of about \unit{36}{\ampere} is defined by the resistance R$_{b}$. The average delay between the laser beam and the discharge onset is \unit{46.0}{\nano\second} with a jitter of \unit{5.7}{\nano\second}.


\subsection{Electron density}

\begin{figure}[!ht]
\begin{center}
\includegraphics[width = 7.5cm]{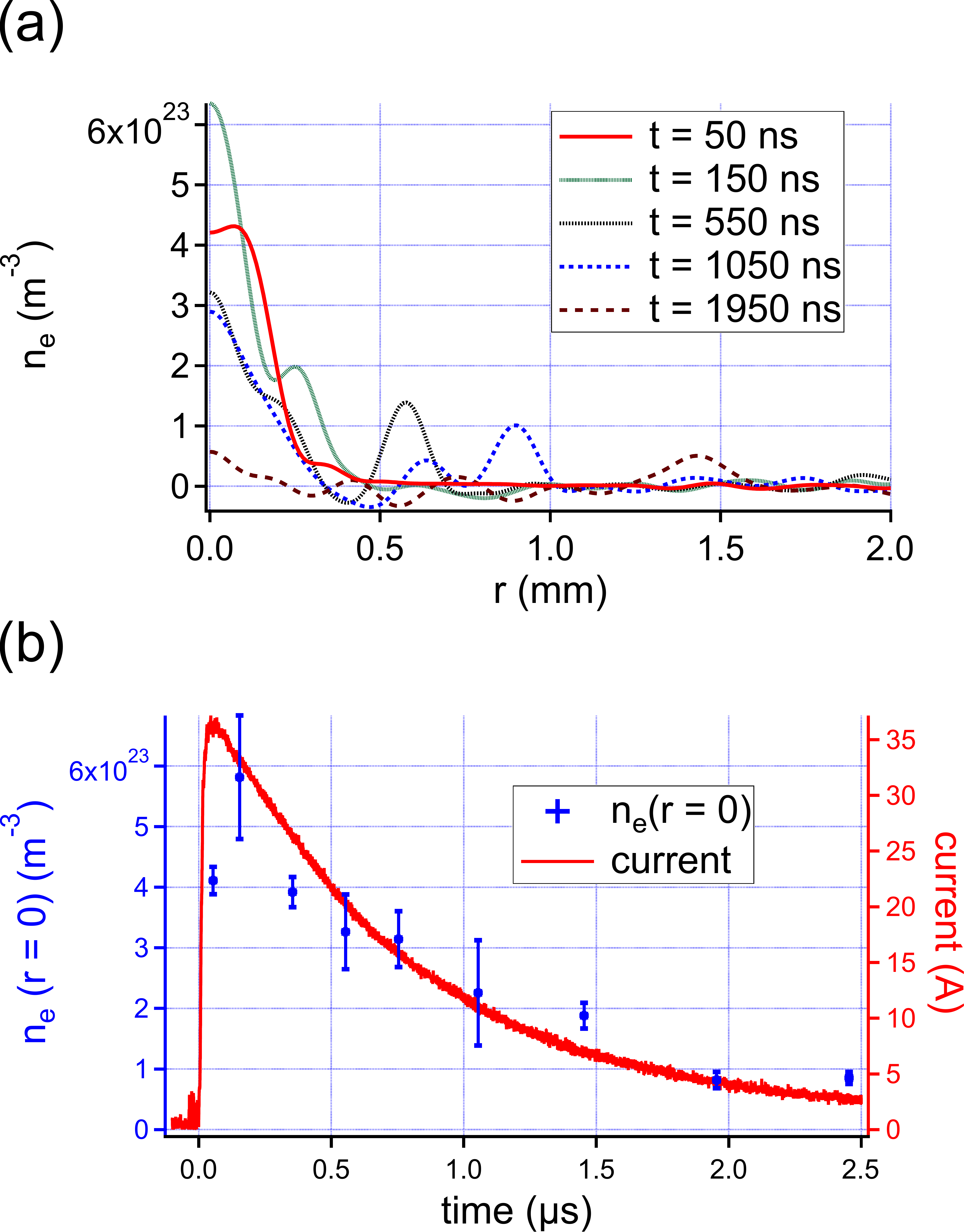}
\end{center}
\caption{(a): examples of electron density profiles for various probing times. (b): on-axis electron densities encountered during the discharge evolution (blue markers), and typical discharge current curve (red solid line).}
\label{fig_electron_density}
\end{figure}

Examples of radial profiles for electron density are given in figure \ref{fig_electron_density}-(a). These profiles are taken midway of the discharge gap and averaged over 20 rows (i. e. \unit{210}{\micro\metre} along $z$). They are characterized by a central main peak, the amplitude of which is at most $\unit{6.8\times10^{23}}{\rpcubic\metre}$ and with a FWHM of $\sim \unit{100}{\micro\metre}$, flanked by a smaller secondary peak with a comparable width. As time goes on, both peaks decrease in amplitude, with the lateral peak flattening much faster and propagating outwards with an average speed of about \unit{700}{\metre\cdot\rp\second}, while keeping approximately a constant width. After \unit{2.5}{\micro\second}, that is almost 3 times the R$_{b}$C$_{d}$ constant for our circuit, it is still possible to record a peak density around $\unit{10^{23}}{\rpcubic\metre}$. Past this point, $n_{e}$ drops below our detection limit.

Figure \ref{fig_electron_density}-(b) displays the evolution of on-axis electron density and of the discharge current with time. Each data point represents an average taken over five different shots, and the corresponding error bar corresponds to plus/minus one standard deviation. On this figure, it is possible to witness the decreasing trend of $n_{e}$ in time, which loosely follows the exponential decrease of the current. This behavior can be easily explained by the fact that the central density peak does not widen significantly in time, as shown in figure \ref{fig_electron_density}-(a). As a consequence, maximum electron density should indeed follow the current time evolution.

The occurrence of the electronic side peak propagating outwards is linked to the formation of a shock wave resulting from energy deposition at the center of the channel. Even though this phenomenon is much more flagrant when studying neutral density profiles (as it will be seen in the next section), it appears that a significant portion of free electrons is carried away by this hydrodynamic wave.

The maximum encountered electron density can be compared to an estimated value using the discharge current following:
\begin{equation}
I = \iint_{S}\vv{j}\cdot\mathrm{d}\vv{S} \sim en_{e}(r=0)\mu E \pi R^{2},
\end{equation}
where $\mu = e/m_{e}\nu_{c}$ is the electron mobility, $E$ the electric field amplitude, and $2R$ the FWHM of the conducting channel. Using $E \sim \unit{15}{\kilo\volt\cdot\centi\rp\metre}$ and $R \sim \unit{100}{\micro\metre}$, we find:
\begin{equation}
n_{e,max}(r = 0) \sim \unit{6\times10^{23}}{\rpcubic\metre},
\end{equation}
which is indeed in very good agreement with the measured value of $\unit{6.8\times10^{23}}{\rpcubic\metre}$.

\subsection{Neutral density}

\begin{figure}[!ht]
\begin{center}
\includegraphics[width = 7.9cm]{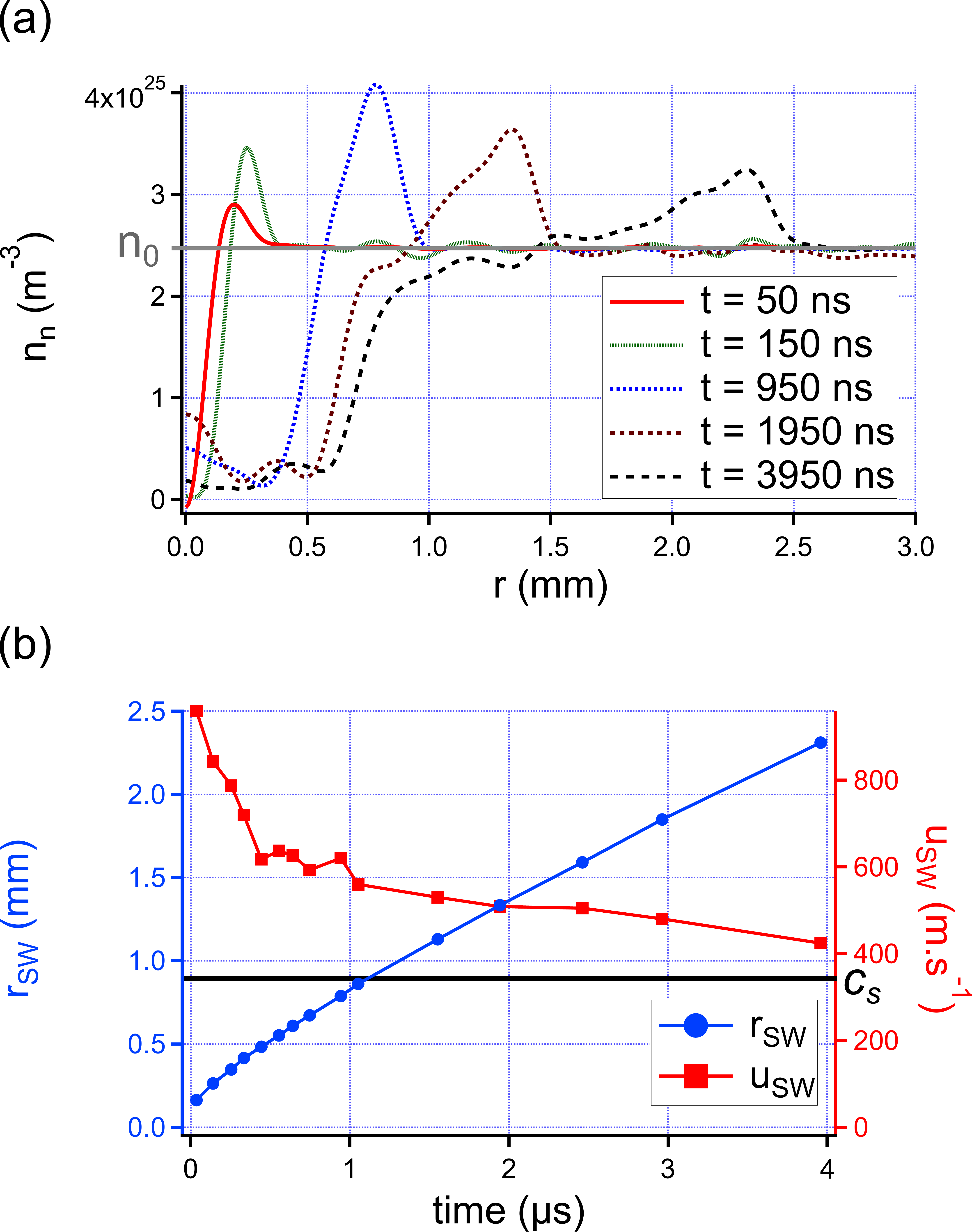}
\end{center}
\caption{(a): measured neutral density profiles for various probing times. (b): radial position of the shock wave, $r_{SW}$ (blue circles) and corresponding speed $u_{SW}$ (red squares). The speed of sound $c_{s}$ is given for convenience.}
\label{fig_neutral_density}
\end{figure}

As seen in figure \ref{fig_neutral_density}-(a), neutral density profiles are characterized by a central, low-density region where $n_{n}$ is typically less than 10 \% of the reference neutral density $n_{0}$, surrounded by a high-density ring.

With time, the central density hole widens while the high-density peak increases in amplitude at first, and then decreases and enlarges. After a few $\upmu$s, the hole FWHM stabilizes around \unit{1.5}{\milli\metre}. This behavior is typical of the formation of a blast wave following an important, very localized energy deposition in the medium, with the formation of an outward-propagating shock wave.

Figure \ref{fig_neutral_density}-(b) represents the time evolution of the shock wave radius, $r_{SW}$, taken as the radial position where neutral density is maximum. Using this data, we can then estimate the shock wave velocity $u_{SW} = \mathrm{d}r_{SW}/\mathrm{d}t$. As seen in the same figure, the velocity remains well above the speed of sound in air in standard temperature and pressure conditions, $c_{s} = \unit{343}{\metre\cdot\rp\second}$, up to the maximum probed time, beyond which measurements cannot be reliably done because $r_{SW}$ exceeds the field of view of cameras. Starting at nearly \unit{1}{\kilo\metre\cdot\rp\second}, $u_{SW}$ falls abruptly during the first \unit{500}{\nano\second} then gently decreases to about \unit{400}{\metre\cdot\rp\second} after \unit{4}{\micro\second}.


\subsection{Study of probe beam deflection by the plasma}

\begin{figure}[!ht]
\begin{center}
\includegraphics[width = 7.5cm]{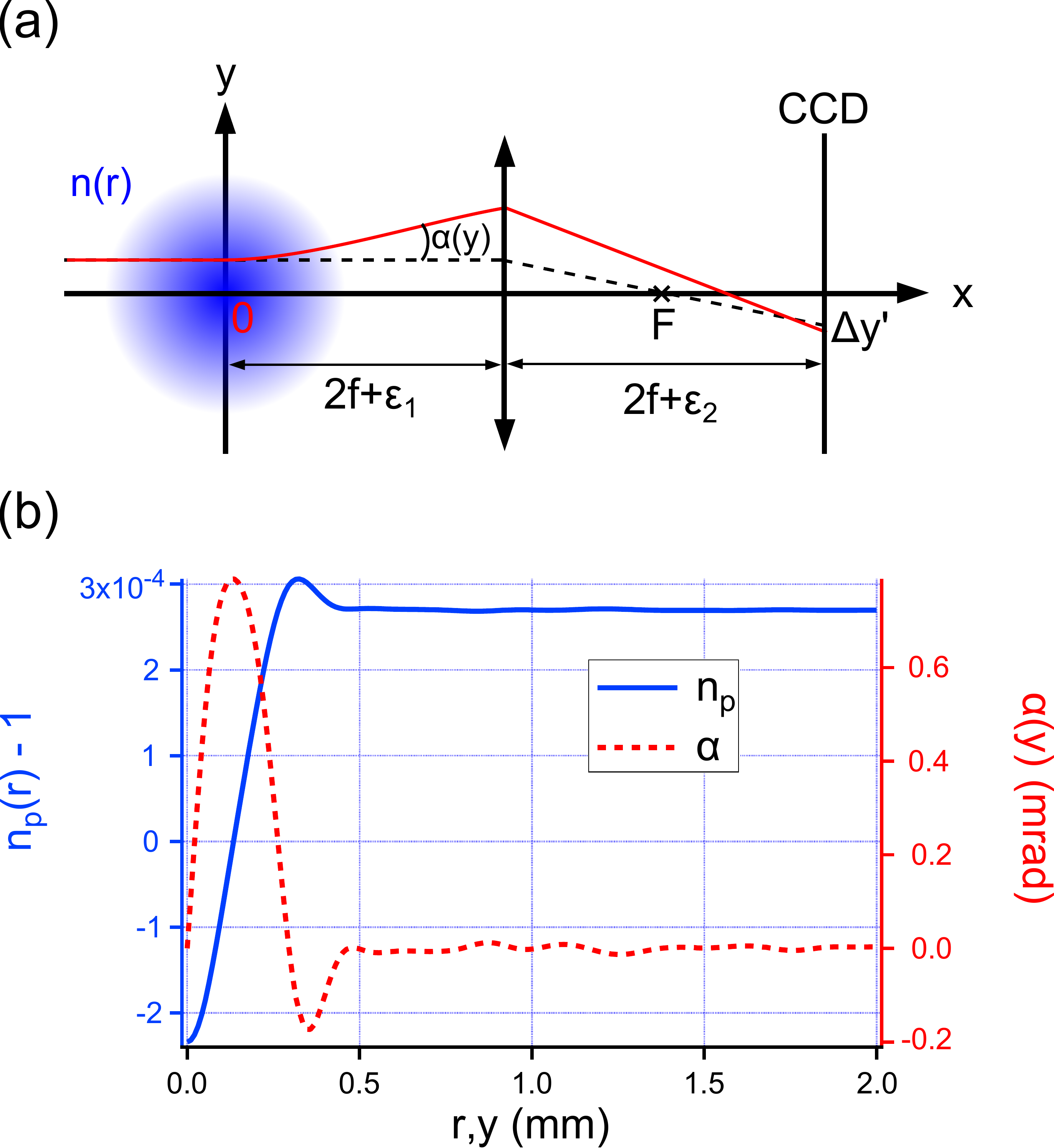}
\end{center}
\caption{(a): schema for probe beam deflection scenario. (b): plasma index at $t = \unit{150}{\nano\second}$ and $\lambda = \unit{1064}{\nano\metre}$ (blue solid line) and corresponding deflection angle (red dashed line).}
\label{fig_beam_deflection}
\end{figure}

No interferometric measurement of a nonhomogeneous refractive index can be rigorously made without checking that probe beam deflection due to index gradients is negligible. In our case, this verification is done at times when index gradients are the most pronounced ($t \sim \unit{150}{\nano\second}$) using index profiles measured at \unit{1064}{\nano\metre}, since this effect is more important at longer wavelengths. The first task to perform is to evaluate the profile for deflection angles, $\alpha(y)$ (cf. figure \ref{fig_beam_deflection}-(a)). In the framework of geometrical optics, this profile can be calculated by solving the following equation:
\begin{equation}
\alpha(y) = \pi-2y(1+\beta) \int_{r_{0}}^{+\infty}\frac{\mathrm{d}r}{r\sqrt{r^{2}n_{p}^{2}-y^{2}(1+\beta)^{2}}},
\end{equation}
where $r_{0}$ is the radius solution of the equation $r_{0}n_{p}(r_{0}) = y(1+\beta)$ \cite{Kogelschatz1972}. Using a refractive index profile recorded at $\lambda = \unit{1064}{\nano\metre}$ and $t = \unit{150}{\nano\second}$, we find the maximum deviation angle to be less than \unit{0.8}{\milli\radian} (figure \ref{fig_beam_deflection}-(b)).

We can also evaluate the position difference $\Delta y'$ on the CCD array between the deflected beam and an hypothetical undeviating beam (respectively the red solid beam and the black dashed beam in figure \ref{fig_beam_deflection}-(a)), given by:
\begin{equation}
\Delta y' = \left(\epsilon_{1}+\epsilon_{2}+\frac{\epsilon_{1}\epsilon_{2}}{f}\right)\tan(\alpha(y)),
\label{eq_ccd_deviation}
\end{equation}
where $2f+\epsilon_{1}$ corresponds to the distance between the imaging lens and the point where the two beams start to separate, and $2f+\epsilon_{2}$ to the distance between the lens and the CCD array. $\epsilon_{1}$ originates from any misplacement of the imaging lens with respect to the plasma , but also from the spatial extension of the plasma, while $\epsilon_{2}$ arises solely from the displacement of the CCD array with respect to its intended position.

For $\Delta y'$ to be as large as one pixel size, the left-hand term in equation \eqref{eq_ccd_deviation} must be on the order of \unit{1}{\centi\metre}. As the plasma has a typical transverse size of \unit{100}{\micro\metre}, the only significant contribution in $\epsilon_{1}$ comes from an eventual misplacement of the imaging lens, much like $\epsilon_{2}$. As a consequence, even a slight displacement of the imaging lens on the order of a few mm cannot result in a detectable influence of the probe deflection on experimental results. We therefore conclude that probe deflection effects can be completely neglected in our experiment.

\section{Summary}

In this Article, we presented a sensitive two-color interferometer that can be used as a plasma diagnostic for simultaneous measurements of free electron density and neutral density with a $\sim \unit{10}{\micro\metre}$ spatial resolution and $\sim \unit{10}{\nano\second}$ time resolution. Interferogram treatment starts with a phase extraction algorithm based on 1D continuous wavelet transform using a Morlet wavelet. Recovered phase is unwrapped by means of a non-continuous path phase unwrapping routine, and phase shift is estimated by subtracting adequate phase reference, allowing to recover line-integrated electron and neutral densities. Finally, a Fourier-Hankel algorithm for Abel inversion of integrated profiles is used. After image treatment, we estimate the phase sensitivity of the interferometer to be \unit{30}{\milli\radian} at the shorter wavelength, giving a resolution of $\sim\unit{4\times10^{22}}{\rpcubic\metre}$ for free electron density and $\unit{10^{24}}{\rpcubic\metre}$ for neutral density with a typical size $\sim \unit{200}{\micro\metre}$ for the studied plasma.

We tested our diagnostic on low-current, cm-scale sparks triggered by means of laser filamentation. We were able to realize space and time resolved measurement of electron and neutral densities and to follow the evolution of corresponding spatial profiles in time until the channel expansion reached the boundary of the field of view of the cameras.

\acknowledgments{This work was funded by the French Direction Générale de l'Armement (Grant n\textsuperscript{o} 2013.95.0901). We also want to thank Florian Mollica for fruitful discussions.}

\bibliographystyle{aipnum4-1}
\bibliography{biblio}

\end{document}